# Investigation of chiral density wave mean field Hamiltonian

*Partha Goswami\**

*Dept. of Physics, D.B.College(University of Delhi), Kalkaji,Delhi-110019,India*

**Abstract.** We start with the chiral(d+id)density wave(CDW)mean field Hamiltonian in momentum space for the pseudo-gapped state of YBCO in the absence of magnetic field, including the momentum conserving inter-layer tunneling matrix elements in the Hamiltonian for the tetragonal case, with the principal component of the order parameter corresponding to $id_{x^2-y^2}$ and the secondary one to the $d_{xy}$. With the eigenvalues and the eigenvectors of the Hamiltonian matrix we write down expression for Berry curvature(BC). Only two BC peaks of same magnitude and opposite signs are obtained at $(\pm\pi/2, \pm\pi/2)$ – the nodal points of the reconstructed Fermi surface (RFS) for the pure id-density wave case. We establish a relation between thermodynamic potential of the system and certain spectral functions for the CDW state. This yields an expression for entropy in closed form. We show that the transition to the CDW state is a first order one and the entropy per unit cell increases in this state. We also investigate anomalous Nernst effect theoretically.

Tel.: +91129 243 9099; fax: +9111 2644 9396. E-mail address: physicsgoswami@gmail.com

## 1. Introduction

The 'pseudo-gap'(PG) behavior of the cuprate high-temperature superconductors in the under-doped regime has been explained by Chakravarty et al.[1,2,3,4] by making a single assumption that $id_{x^2-y^2}$ density- wave (DDW) state corresponds to PG. The DDW order parameter (OP) is purely imaginary in the momentum space for the ordering wave vector $\mathbf{Q}=(\pm\pi,\pm\pi)$(commensurate magnetic fluctuations). The DDW order corresponds to spontaneous currents along the bonds of a square lattice for the ordering wave vector $(\pm\pi,\pm\pi)$. Although the currents violate the microscopic time reversal symmetry (MTRS), there is no macroscopic violation. This happens as the DDW state preserves the combined effect of MTRS and translation by a lattice spacing. The net result is that the staggered magnetic flux produced by these currents is zero on the macroscopic scale. Since the commensurate DDW order also doubles the unit cell of the real space lattice, the conventional Brillouin zone (BZ) in the reciprocal space is halved, or reduced, which is known as the reduced Brillouin zone (RBZ). This unit cell doubling plays a crucial role in an analysis of spectral functions involving DDW order. Within the framework of these ideas, the recent experimental observation [5,6,7] of Shubnikov de Haas oscillation (SdHO) in longitudinal conductance of $YBa_2Cu_3O_{7-\delta}$ (YBCO) close to 0 K has been explained satisfactorily[4].

The observation of a non-zero polar Kerr effect(PKE),which violates the macroscopic time reversal symmetry(TRS), in the under-doped YBCO[8] at temperatures $T<T_s$ (p) (where p denotes hole concentrations) close to the pseudo-gap temperature T* (p) strongly indicates that the interconnectedness of the pseudo-gap in the cuprates and the macroscopic TRS breaking. The temperature $T_s(p)$ drops rapidly with increasing hole concentration. Now the magnitude of the Kerr rotation in YBCO is found to be smaller by ~ 4 orders of magnitude than that observed in other itinerant ferromagnetic oxides by the authors in ref.[8]. This suggests that either the PKE not directly measuring the principal OP characterizing the pseudo-gap phase in YBCO, or measuring a very small "ferromagnetic-like" component of OP. Das Sarma et al.[9] have been successful to account for the observed PKE by the inclusion of $d_{xy}$ component in the principal $id_{x2-y2}$ component of the total OP(chiral (d+id) density-wave(CDW) state). The former leads to the staggered modulation of the diagonal electron tunneling between the next-nearest neighbor lattice sites. Such staggered modulation breaks the symmetry between the plaquettes with positive and negative circulation and leads to the violation of the macroscopic TRS. Due to this, the ordered state acquires a Berry curvature (BC)[10] which has peaks at $(\pm\pi/2, \pm\pi/2)$ –the nodal points, corresponding to the gap-less excitations, of the reconstructed Fermi surface(RFS) for the pure DDW order.

The aforementioned experimental observations [5,6,7,8] indicate that the identification of PG behavior with the d+id (chiral) density wave order, with the principal component of the order parameter corresponding to $id_{x2-y2}$ and the secondary one to the $d_{xy}$, is more appropriate. In the present communication, we start with a Hamiltonian $H_{d+id}$ in momentum space; the pure DDW order corresponds to a special case. Since the Hamiltonian is to describe YBCO, it includes the momentum conserving inter-layer tunneling matrix elements for the tetragonal structure. Our first aim is to derive expressions for BC and entropy in the absence of magnetic field. The latter is useful for obtaining finite temperature thermodynamics of the system. We show that the transition to CDW state is a first order one and the entropy per unit cell increases in this state. We also investigate the spontaneous(anomalous)Nernst effect (SNE) with the expressions for BC and entropy derived.

The paper is organized as follows: In section 2 we start with the d+id density wave mean field Hamiltonian ($H_{d+id}$) in momentum space in the absence of magnetic field. We calculate the eigenvalues and the eigenvectors of the Hamiltonian matrix. With the aid of the latter we write down expression for BC. In section 3, we establish first a relation between thermodynamic potential of the system and certain spectral functions for the CDW state. This yields an expression for entropy in closed form. The expression for entropy together with that of BC allows us to discuss anomalous Nernst effect in section 4.

## 2. Chiral d+id density-wave Hamiltonian

In the second-quantized notation, the Hamiltonian (with index j = (1,2) below corresponding to bi-layer system) in the Chiral d+id density-wave state can be expressed as $H_{d+id} = \sum_{k\sigma,i=1,2} \Phi^{(j)\dagger}_{k,\sigma} E(k) \Phi^{(j)}_{k,\sigma}$ where $\Phi^{(j)\dagger}_{k,\sigma} = (d^{\dagger(1)}_{k,\sigma}\ d^{\dagger(1)}_{k+Q,\sigma}\ d^{(2)\dagger}_{k,\sigma}\ d^{\dagger\ (2)}_{k+Q,\sigma})$ and

$$E(k) = \begin{pmatrix} \varepsilon_k & D_k \exp(i\theta_k) & t_k & 0 \\ D_k \exp(-i\theta_k) & \varepsilon_{k+Q} & 0 & t_{k+Q} \\ t_k & 0 & \varepsilon_k & D_k \exp(i\theta_k) \\ 0 & t_{k+Q} & D_k \exp(-i\theta_k) & \varepsilon_{k+Q} \end{pmatrix}. \quad (1)$$

Here the chiral order parameter(COP) $D_k \exp(i\theta_k)$ is expressed in terms of the order parameter(OP) amplitudes of the $d_{xy}$ component, $\chi_k = -\chi_0 \sin(k_x a) \sin(k_y a)$, and that of the principal $id_{x2-y2}$ component, $\Delta_k = (\Delta_0(T)/2)(\cos k_x a - \cos k_y a)$, as $D_k = (\chi_k^2 + \Delta_k^2)^{1/2}$ with $(-\chi_k) = D_k \cos\theta_k$ and $\Delta_k = D_k \sin\theta_k$. In the pure DDW state $\chi_k = 0$ and therefore $\theta_k = \pi/2$ is a constant. Also, as we shall see below, an important deviation from the pure DDW state is that, with COP, gap-less excitations shall be unavailable on the reconstructed Fermi surface. The normal state tight-binding energy dispersion $\varepsilon_k = \varepsilon_k^{(1)} + \varepsilon_k^{(2)} + \varepsilon_k^{(3)} - \mu$, $\varepsilon_k^{(1)} = -2t(c_x + c_y)$, $\varepsilon_k^{(2)} = 4t' c_x c_y$, $\varepsilon_k^{(3)} = -2t''(c'_x + c'_y)$, $c_j = \cos k_j a$, $c'_j = \cos 2k_j a$ ($j = x,y$), and 'a' is the lattice constant. The quantity $\varepsilon_k$ involves $t, t', t''$ which are the hopping elements between nearest, next-nearest(NN) and NNN neighbors, respectively, and $\mu$ is the chemical potential. The energy $\varepsilon^{(1)}(\mathbf{k})$ satisfies the perfect nesting condition $\varepsilon^{(1)}(\mathbf{k+Q}) = -\varepsilon^{(1)}(\mathbf{k})$ with ordering wave vector $\mathbf{Q} = (\pm\pi, \pm\pi)$. The effect of bi-layer splitting in YBa$_2$Cu$_3$O$_{7-\delta}$ (YBCO) is given by the parametrization in terms of a momentum conserving tunneling matrix element which for the tetragonal structure corresponds to $t_k = (t_0/4)(\cos k_x a - \cos k_y a)^2$. The energy eigenvalues of the matrix in Eq.(1) (with $j = \pm 1$ ($j = +1$ corresponds to the upper branch(U) and $j = -1$ to the lower branch(L) ), and $\nu = \pm 1$) are $E^{(j,\nu)}(k) = [\varepsilon_k^U + jw_k + \nu t_k]$ where $\varepsilon_k^U = (\varepsilon_k + \varepsilon_{k+Q})/2$, $w_k = [(\varepsilon_k^L)^2 + D^2_k]^{1/2}$ and $\varepsilon_k^L = (\varepsilon_k - \varepsilon_{k+Q})/2$. The corresponding eigenvectors are given by

$$\Psi^{(U,\nu)}(k) = \begin{bmatrix} u_k \exp(i\theta_k/2) \\ v_k \exp(-i\theta_k/2) \\ \nu u_k \exp(i\theta_k/2) \\ \nu v_k \exp(-i\theta_k/2) \end{bmatrix}, \quad \Psi^{(L,\nu)}(k) = \begin{bmatrix} v_k \exp(i\theta_k/2) \\ -u_k \exp(-i\theta_k/2) \\ \nu v_k \exp(i\theta_k/2) \\ -\nu u_k \exp(-i\theta_k/2) \end{bmatrix}. \quad (2)$$

The $(u_k^2, v_k^2)$ are coherence factors given by $u_k^2$, $v_k^2 = (1/2)(1 \pm (\varepsilon_k^L/w_k))$. Therefore, though $t_k$ causes band-splitting, the coherence factors (and $\theta_k$) are independent of $t_k$. As we shall see below, for this reason, the Berry curvatures(BC)[10] do not depend on $t_k$.

We now outline the calculation of BC, $\Omega(\mathbf{k})$, for the system under consideration. The BC effect, which met with much success in explaining spontaneous Nernst effect(SNE) in ferro-magnets [11,12], will be taken into account to explain SNE here in the absence of magnetic field. Another reason for the emphasis on BC is that for the investigation of the magnetic field(**B**) dependent phenomena, such as the quantum oscillations in longitudinal conductivity, shown by the system whenever we need to replace the sum over physical momenta **k** (gauge-invariant but non-canonical) by an integral over a region of **k**-space, the Berry-phase corrected result[11] is $\sum_\mathbf{k} \to \int d\mathbf{k}\, (1 + (e\mathbf{B}\cdot\Omega(\mathbf{k})/\hbar))$. We shall see that for some isolated points in **k**-space, the quantity $(e\mathbf{B}\cdot\Omega(\mathbf{k})/\hbar))$ is comparable to unity for a magnetic field B ~ 1 Tesla. Effectively, these points contribute to the Berry-phase correction which is the momentum integral of BC, $N = \int_{BZ} d\mathbf{k}\, (e\mathbf{B}\cdot\Omega(\mathbf{k})/\hbar)$, known as the Chern number. The formal expression for the BCs, for the problem at hand, may be written down as $\Omega^{(j,v)}(\mathbf{k}) = \nabla_\mathbf{k} \times \langle \Psi^{\dagger(j,v)}(\mathbf{k}) | i\nabla_\mathbf{k} \Psi^{(j,v)}(\mathbf{k})\rangle$. The gradients of $\Psi^{(j,v)}(\mathbf{k})$ are directly obtainable from Eq.(3). In fact, since here the momentum **k** is confined to ($k_x$, $k_y$), only the transverse components $\Omega^{(j,v)}(k_x, k_y)$ of $\Omega^{(j,v)}(\mathbf{k})$ are non-zero. We find that

$$\Omega^{(j)}(k_x, k_y) = \pm\, [\nabla_k (u_k^2 - v_k^2) \times \nabla_k \theta_k]\big|_{\text{z-component}}$$

$$= \pm\, w_k^{-3}[\chi_k\{-\Delta_k, \varepsilon_k^L\} + \Delta_k\{\varepsilon_k^L, -\chi_k\} - \varepsilon_k^L\{-\chi_k, -\Delta_k\}]$$

$$= \pm\, ta^2 w_k^{-3} \chi_0 \Delta_0 (\sin^2 k_y a + \sin^2 k_x a \cos^2 k_y a) \qquad (3)$$

where $\{a_k, b_k\} = (\partial a_k/\partial k_x)(\partial b_k/\partial k_y) - (\partial b_k/\partial k_x)(\partial a_k/\partial k_y)$. In Fig.1 we have plotted the Berry curvatures $\Omega^{(1)}(k_x, k_y)$ on the Brillouin zone at 10% hole doping for which the parameter value $(\Delta_0/t) = 0.3200$. We also assume $(\chi_0/t) = 0.0160$. We find that $\Omega^{(j)}(k_x, k_y)$ peak dramatically at $(\pm\pi/2, \pm\pi/2)$ –the nodal points, corresponding to the gap-less excitations, of the reconstructed Fermi surface(RFS) for the pure DDW order[3,4]. In the absence of impurity scattering etc., the lifetime of these peaks is infinite as the Hamiltonian is quadratic in fields and therefore exactly diagonalizable. As a consequence, the eigenstates given by Eq.(2) are stationary states with infinite lifetime (and these stationary states have been used to extract the peaks). Elsewhere on the RFS, including the anti-nodal points $\{(\pm\pi, 0), (0,\pm\pi)\}$, however, $\Omega^{(j)}(k_x, k_y)$ remain almost zero. The precise values of $\Omega^{(j)}(\mathbf{k}a=(\pm\pi/2, \pm\pi/2))/a^2$ are $\pm 9.9609\times10^{3}$. For YBCO, the lattice constant a = 3.82 $A^0$. With increase in the hole concentration $\delta$, the reduced Brilliouin zone boundary(RBZ) boundary (see Fig.2 where we have plotted BC, $\Omega^{(1)}$, at 14% hole concentration) gets outlined clearly. For the variation of $\Delta_0(\delta)$ as a function of $\delta$, we have used the approximate formula similar to that of Loram et al.[13]. The BC peak value is, however, found to be doping-independent, since at $\mathbf{k}a=(\pm\pi/2, \pm\pi/2)$ the gap parameter $\Delta_k(\delta)$ does not contribute to $\Omega^{(j)}(k_x, k_y)$. We note that, though the energy eigenvalues and the corresponding eigenstates bear the signature of the in-plane momentum conserving tunneling process (for the tetragonal case), the BC's are completely free from this effect(as the coherence factors and $\theta_k$ are independent of $t_k$).

## 3. Thermodynamic potential and entropy

In this section we establish first a relation between thermodynamic potential $\Omega_{d+id} = -(1/\beta) \ln \text{Tr} \exp(-\beta H_{d+id})$ of the system and certain spectral functions. This yields an expression for entropy in closed form. The relation to be obtained is useful for deriving thermodynamics of the system in the pseudo-gapped state. The methodology followed is similar to that of Kadanoff and Baym[14]. About five decades ago these authors had established a formula relating thermodynamic average of a model Hamiltonian to a spectral weight function for an interacting Bose system in the normal phase. For the purpose stated, it is convenient to define a new thermodynamic potential $\Omega_{d+id}(\lambda)$ in terms of the Hamiltonian $H_{d+id}(\lambda) = \lambda H_{d+id}$ where $\lambda$ is a variable. One can write $\Omega_{d+id}(\lambda) - \Omega(0) = \int (d\lambda/\lambda) \langle H_{d+id}(\lambda) \rangle_\lambda$ where $\Omega(0)$ is an integration constant and the angular brackets $\langle \ldots \rangle_\lambda$ denote thermodynamic average calculated with $H_{d+id}(\lambda)$. The system under consideration corresponds to $\Omega_{d+id}(\lambda = 1)$. Obviously, the task now boils down to establishing relation between the average $\langle H_{d+id}(\lambda) \rangle_\lambda$ and spectral weight functions. The spectral functions are given by

$$A^{(j)}_\lambda(k,\sigma,\omega) = i[G^{(j)}_\lambda(k_1,\sigma,\omega_n)|_{i\omega_n = \omega + i0+} - G^{(j)}_\lambda(k_1,\sigma,\omega_n)|_{i\omega_n = \omega - i0+}], \quad (4)$$

where $G^{(j)}_\lambda(k,\sigma, k',\sigma',\omega_n) = \int_0^\beta d\tau\, e^{i\omega_n\tau} G^{(j)}_\lambda(k,\sigma,\tau; k',\sigma',0)$, $G^{(j)}_\lambda(k,\sigma,\tau; k',\sigma', \tau') = -\langle T\{d^{(j)}_{k,\sigma}(\tau) d^{(j)\dagger}_{k',\sigma'}(\tau')\}\rangle_\lambda$, $d^{(j)}_{k,\sigma}(\tau) = \exp(H_{d+id}(\lambda)\tau) d^{(j)}_{k,\sigma} \exp(-H_{d+id}(\lambda)\tau)$, $\omega_n = [(2n+1)\pi/\beta]$ with $n = 0, \pm 1, \pm 2, \ldots \ldots$, and T is the time-ordering operator which arranges other operators from right to left in the ascending order of imaginary time $\tau$. The Lehmann representation (LR) of $A^{(j)}_\lambda(k,\sigma,\omega)$ can be obtained easily with that of $G^{(j)}_\lambda(k,\sigma,\omega_n)$. For the former, we find

$$A^{(j)}_\lambda(k,\sigma,\omega) = 2\pi\exp(\beta\Omega_{d+id}(\lambda)) \sum_{mn} e^{-\beta H_m(\lambda)} \langle m|d^{(j)\dagger}_{k,\sigma}|n\rangle \langle n|d^{(j)}_{k,\sigma}|m\rangle$$
$$\times (e^{\beta\omega} + 1)\, \delta(\omega + H_n(\lambda) - H_m(\lambda)). \quad (5)$$

Here $|m\rangle$ is an exact eigen state of $H_{d+id}(\lambda)$ and $\{H_{d+id}(\lambda)|m\rangle = H_m(\lambda)|m\rangle\}$. It is convenient to introduce the functions $f^{(j)}_\lambda(k,\sigma,\omega)$ and $f^{(j)}_\lambda(k+Q,\sigma,\omega)$, where

$$f^{(j)}_\lambda(k,\sigma,\omega) \equiv [-i \iint dt\, dt'\, e^{i\omega(t-t')} \langle d^{(j)\dagger}_{k,\sigma}(t') d^{(j)}_{k,\sigma}(t)\rangle_\lambda \Theta(t-t')] \quad (6)$$

and $\Theta(t)$ is the unit step-function given by $\Theta(t) = i \int_{-\infty}^{+\infty}(d\omega/2\pi)(e^{-i\omega t}/(\omega + i0^+))$, for our purpose. Here $0^+$ tends towards zero from positive values. The LR of $f^{(j)}_\lambda(k,\sigma,\omega)$ and $f^{(j)}_\lambda(k+Q,\sigma,\omega)$ can be obtained easily using this integral representation of $\Theta(t)$ and the identity $(x \pm i\,0^+)^{-1} = [P(x^{-1}) \pm (1/i)\,\pi\,\delta(x)]$ valid for real $\omega$, where P represents a Cauchy's principal value. We find

$$\text{Im } f^{(j)}_\lambda(k,\sigma,\omega) = -\pi\exp(\beta\Omega_{d+id}(\lambda))\sum_{mn} e^{-\beta H_m(\lambda)} \langle m|d^{(j)\dagger}_{k,\sigma}|n\rangle\langle n|d^{(j)}_{k,\sigma}|m\rangle$$

$$\times \delta(\omega + H_n(\lambda) - H_m(\lambda)), \quad (7)$$

and

$$\text{Re } f^{(j)}_\lambda(k_1,\sigma,\omega) = -P\int_{-\infty}^{+\infty}(d\omega'/\pi)\{\text{Im } f^{(j)}_\lambda(k_1,\sigma,\omega)/(\omega-\omega')\}. \quad (8)$$

Upon comparing (7) with (6) we obtain $\text{Im } f^{(j)}_\lambda(k,\sigma,\omega) = (-1/2)(e^{\beta\omega}+1)^{-1} A^{(j)}_\lambda(k,\sigma,\omega)$ and a similar one: $\text{Im } f^{(j)}_\lambda(k+Q,\sigma,\omega) = (-1/2)(e^{\beta\omega}+1)^{-1} A^{(j)}_\lambda(k+Q,\sigma,\omega)$. In view of this result and Eq.(8) one can write

$$f^{(j)}_\lambda(k,\sigma,\omega) = [-i\iint dt\,dt'\int_{-\infty}^{+\infty}(d\omega'/2\pi)e^{i(\omega-\omega')(t-t')}(e^{\beta\omega'}+1)^{-1} A^{(j)}_\lambda(k,\sigma,\omega')\Theta(t-t')] \quad (9)$$

and a similar one for which $f^{(j)}_\lambda(k+Q,\sigma,\omega)$. Upon comparing (6) with (9) one may write

$$\langle d^{(j)\dagger}_{k,\sigma}(t')d^{(j)}_{k,\sigma}(t)\rangle_\lambda = \int_{-\infty}^{+\infty}(d\omega'/2\pi)e^{-i\omega'(t-t')}(e^{\beta\omega'}+1)^{-1} A^{(j)}_\lambda(k,\sigma,\omega'),$$

$$\langle d^{(j)\dagger}_{k+Q,\sigma}(t')d^{(j)}_{k+Q,\sigma}(t)\rangle_\lambda = \int_{-\infty}^{+\infty}(d\omega'/2\pi)e^{-i\omega'(t-t')}(e^{\beta\omega'}+1)^{-1} A^{(j)}_\lambda(k+Q,\sigma,\omega'). \quad (10)$$

We now wish to express $\langle H_{d+id}(\lambda)\rangle_\lambda$ in terms of the averages $\langle d^{(j)\dagger}_{k,\sigma}(t')d^{(j)}_{k,\sigma}(t)\rangle_\lambda$ and $\langle d^{(j)\dagger}_{k+Q,\sigma}(t')d^{(j)}_{k+Q,\sigma}(t)\rangle_\lambda$ in (10) to obtain an expression for the thermodynamic potential in terms of spectral functions.

For the purpose stated above we set up equations for the operators $\acute{O}(t) = d^{(j)}_{k,\sigma}(t)$, $d^{(j)}_{k+Q,\sigma}(t)$, etc. using the equations of motion(in units such that $\hbar = 1$) $i(\partial/\partial t)\acute{O}(t) = [\acute{O}(t), H_{d+id}(\lambda)]$, where $\acute{O}(t) = \exp(iH_{d+id}(\lambda)t)\,\acute{O}\,\exp(-iH_{d+id}(\lambda)t)$. With the help of these equations we shall then express $\langle H_{d+id}(\lambda)\rangle_\lambda$ in terms of the averages $\langle d^{(j)\dagger}_{k,\sigma}(t')d^{(j)}_{k,\sigma}(t)\rangle_\lambda$ and $\langle d^{(j)\dagger}_{k+Q,\sigma}(t')d^{(j)}_{k+Q,\sigma}(t)\rangle_\lambda$. The equations of motion $i(\partial/\partial t)\acute{O}(t) = [\acute{O}(t), H_{d+id}(\lambda)]$ yield

$$\sum_{j,k,\sigma} \{(i/\lambda)(\partial/\partial t) - (i/\lambda)(\partial/\partial t')\} \langle d^{(j)\dagger}_{k,\sigma}(t') d^{(j)}_{k,\sigma}(t) \rangle_\lambda$$

$$= 2\sum_{j,k,\sigma} \varepsilon_k \langle d^{(j)\dagger}_{k,\sigma}(t') d^{(j)}_{k,\sigma}(t) \rangle_\lambda + \sum_{j,k,\sigma} D_k \exp(i\theta_k) \langle d^{(j)\dagger}_{k,\sigma}(t') d^{(j)}_{k+Q,\sigma}(t) \rangle_\lambda$$

$$+ \sum_{j,k,\sigma} D_k \exp(-i\theta_k) \langle d^{(j)\dagger}_{k+Q,\sigma}(t') d^{(j)}_{k,\sigma}(t) \rangle_\lambda + \sum_{k,\sigma,j(j\neq l)} t_k \langle d^{(j)\dagger}_{k,\sigma}(t') d^{(l)}_{k,\sigma}(t) \rangle_\lambda \quad (11)$$

and

$$\sum_{j,k,\sigma} \{(i/\lambda)(\partial/\partial t) - (i/\lambda)(\partial/\partial t')\} \langle d^{(j)\dagger}_{k+Q,\sigma}(t') d^{(j)}_{k+Q,\sigma}(t) \rangle_\lambda$$

$$= 2\sum_{j,k,\sigma} \varepsilon_{k+Q} \langle d^{(j)\dagger}_{k+Q,\sigma}(t') d^{(j)}_{k+Q,\sigma}(t) \rangle_\lambda + \sum_{j,k,\sigma} D_k \exp(i\theta_k) \langle d^{(j)\dagger}_{k,\sigma}(t') d^{(j)}_{k+Q,\sigma}(t) \rangle_\lambda$$

$$+ \sum_{j,k,\sigma} D_k \exp(-i\theta_k) \langle d^{(j)\dagger}_{k+Q,\sigma}(t') d^{(j)}_{k,\sigma}(t) \rangle_\lambda + \sum_{k,\sigma, j(j\neq l)} t_{k+Q} \langle d^{(j)\dagger}_{k+Q,\sigma}(t') d^{(l)}_{k+Q,\sigma}(t) \rangle_\lambda$$

$$(12)$$

which, in turn, lead to

$$\langle H_{d+id}(\lambda) \rangle_\lambda = (1/2) \lim_{t'\to t} \sum_{j,k,\sigma} \{i(\partial/\partial t) - i(\partial/\partial t')\} \{ \langle d^{(j)\dagger}_{k,\sigma}(t') d^{(j)}_{k,\sigma}(t) \rangle_\lambda$$

$$+ \langle d^{(j)\dagger}_{k+Q,\sigma}(t') d^{(j)}_{k+Q,\sigma}(t) \rangle_\lambda \} \quad (13)$$

for the chiral d-density wave state. Thus the relation, between the thermodynamic potential and the spectral functions, sought for is

$$\Omega_{d+id}(\lambda) - \Omega(0) = \int (d\lambda/\lambda) \langle H_{d+id}(\lambda) \rangle_\lambda$$

$$= \int (d\lambda/\lambda) \sum_{j,k,\sigma} \int_{-\infty}^{+\infty} (d\omega/2\pi) \omega (e^{\beta\omega} + 1)^{-1} \{A^{(j)}_\lambda(k,\sigma,\omega) + A^{(j)}_\lambda(k+Q,\sigma,\omega)\}. \quad (14)$$

The straightforward method of calculating these weight functions comprises of setting up equations of motion for the temperature functions $G^{(j)}_\lambda(k,\sigma,\tau; k',\sigma',\tau')$ and then obtain the corresponding Matsubara propagators $G^{(j)}_\lambda(k,\sigma,k',\sigma',\omega_n)$. The weight functions $(A^{(j)}_\lambda(k,\sigma,\omega), A^{(j)}_\lambda(k+Q,\sigma,\omega))$ can then be obtained using Eq.(4).

For the chiral d-density wave state we find that the weight functions $\sum_v A^{(v)}_\lambda(k,\sigma,\omega) = 2\pi\sum_v [u^{(v)2}_k \delta(\omega - \lambda E_k^{(U,v)}) + v^{(v)2}_k \delta(\omega - \lambda E_k^{(L,v)})]$ and $\sum_v A^{(v)}_\lambda(k+Q,\sigma,\omega) = 2\pi\sum_v [v^{(v)2}_k \delta(\omega - \lambda E_k^{(U,v)}) + u^{(v)2}_k \delta(\omega - \lambda E_k^{(L,v)})]$ where $u^{(v)2}_k = (1/4)[1 + (\varepsilon_k^L / w_k)]$ and $v^{(v)2}_k =$

$(1/4)[1-(\varepsilon_k^L/w_k)]$. The single particle excitation spectrum is $E_k^{(U(L),\nu)}=(\varepsilon_k^U+\nu t_k \pm w_k)$ where $\nu=\pm 1$. This eventually yields the expression for the thermodynamic potential per unit cell as a function of $\lambda$: $\Omega_{d+id}(\lambda)-\Omega(0) = \int d\lambda \ (N_s)^{-1} \sum_k [E_k^{(U,\nu)}(1-\tanh(\beta\lambda E_k^{(U,\nu)}/2))+E_k^{(L,\nu)}(1-\tanh(\beta\lambda E_k^{(L,\nu)}/2))]$ where $N_s$ is the number of unit cells and $\beta = (k_B T)^{-1}$. For the real system, in the absence of magnetic field, one can therefore write

$$\Omega_{d+id}(\lambda=1) = \Omega_0 - 2(\beta N_s)^{-1} \sum_{k,r(=U,L),\nu} \{\ln\cosh(\beta E_k^{(r,\nu)}/2)\} \tag{15}$$

where $\Omega_0 = (N_s)^{-1} \sum_{k,r(=U,L),\nu} E_k^{(r,\nu)}$. The dimensionless entropy per unit cell is given by $s_{d+id} = -(\partial \Omega_{d+id}/\partial (k_B T)) = \beta^2 (\partial \Omega_{d+id}/\partial \beta)$. We obtain for the pseudo-gapped phase and the normal (non-pseudo-gapped) phase, respectively,

$$s_{d+id} = (2/N_s) \sum_{k,r(=U,L),\nu} [\ln(1/2) + \ln(1+\exp(-\beta E_k^{(r,\nu)}))$$
$$+ (\beta E_k^{(r,\nu)} + \beta^2 (\partial E_k^{(r,\nu)}/\partial \beta))(\exp(\beta E_k^{(r,\nu)})+1)^{-1}]. \tag{16}$$

and

$$s_{normal} = (2/N_s) \sum_{k,j=1,2} [\ln(1/2) + \ln(1+\exp(-\beta \varepsilon_j(k)))$$
$$+ (\beta \varepsilon_j(k) + \beta^2 (\partial \varepsilon_j(k)/\partial \beta))(\exp(\beta \varepsilon_j(k))+1)^{-1}] \tag{17}$$

where $\varepsilon_1(k) = \varepsilon_k$, and $\varepsilon_2(k) = \varepsilon_{k+Q}$. The specific heat per unit cell can now be formally expressed as $c = -\beta (\partial s/\partial \beta)$. The parameters we choose (see ref.[3,4]) for the numerical calculation given below, at 10% hole doping, are: $t = 0.25$ eV, $t' = 0.4t$, $t'' = 0.0444\ t$, and $t_0 = 0.032\ t$. With these choices of the parameters, the previous authors[3,4] have noted the value of the chemical potential $\mu$ to be $-0.27$ eV. The pseudo-gap temperature $T^* \sim 150$ K. We assume the experimental value $\Delta_0\ (T < T^*) = 0.0825$ eV $= 0.3300\ t$ in the vicinity of $T^*$, $(\chi_0/\Delta_0 (T<T^*))^2 = 0.0025$, and the quantity $t_0 = 0.032\ t$. Note that $T^*$, $\chi_0$ and $\Delta_0$ could be calculated at a given doping level, in principle, setting up equations for the gap functions and the propagators $G^{(j)}(k,\sigma,\tau; k',\sigma',\tau') = -\langle T\{ d^{(j)}_{k,\sigma}(\tau) d^{(j)\dagger}_{k',\sigma'}(\tau')\}\rangle$

( where $d^{(j)}_{k,\sigma}(\tau) = \exp(H_{d+id}\tau)\, d^{(j)}_{k,\sigma}\, \exp(-H_{d+id}\tau)$). The Luttinger sum rule (LSR) relates the doping level $\delta$ to $(N_s)^{-1} \sum_{k,\sigma,v} \langle d^{(v)\dagger}_{k,\sigma} d^{(v)}_{k,\sigma}\rangle$ where the average occupation number

$$f(k,v) = \langle d^{(v)\dagger}_{k,\sigma} d^{(v)}_{k,\sigma}\rangle = [\, u^{(v)2}_k (\exp(\beta E_k^{(U,v)})+1)^{-1} + v^{(v)2}_k (\exp(\beta E_k^{(L,v)})+1)^{-1}\,]. \quad (18)$$

We, however, did not attempt here to solve such equations which are consistent with $H_{d+id} = \sum_{k\sigma, i=1,2} \Phi^{(j)\dagger}_{k,\sigma} E(k)\, \Phi^{(j)}_{k,\sigma}$. We relied on the well-known values given above. In Fig.3(a) we have plotted $f(k,v,T<T^*)$, whereas in Fig.3(b) we have plotted $f(k,v,T>T^*)$, on the Brillouin zone. The former clearly shows the well-separated electron and hole sectors but, in the latter, this distinction is absent due to the disappearance of the gap in the lower and the upper branches of the excitation spectrum, $E_k^{(U(L),v)} = (\varepsilon_k^U + vt_k \pm w_k)$, $v = \pm 1$. The occupation number remains close to unity as is required by the LSR.

We wish to calculate $\Delta s = (s_{d+id}(T<T^*) - s_{normal}(T>T^*))$, for a doping level $\delta$, to ascertain whether the pseudo-gap transition is the first order one or not. In the foregoing discussion, we have assumed the experimental value $\Delta_0(T<T^*) = 0.0825$ eV $= 0.3300$ t in the vicinity of $T^*$ and the quantity $t_0 = 0.032$ t. Since $t_0$ is smaller by $\sim$ one order of magnitude with relative to $\Delta_0(T)$, the effect of the bi-layer splitting will be ignored in the calculation of $\Delta s$ below. In the vicinity of $T^*$, upon treating the gap $D_k^2$ as a small parameter, we see that

$$\sum_{k,r(=U,L),v} \ln(1+\exp(-\beta E_k^{(r,v)})) \approx \sum_{k,j=1,2} \ln(1+\exp(-\beta \varepsilon_j(k)))$$

$$+ \sum_k (\beta D_k^2 / 2\, \varepsilon_k^L)\{(\exp(\beta \varepsilon_{k+Q})+1)^{-1} - (\exp(\beta \varepsilon_k)+1)^{-1}\}$$

$$+ \sum_k (\beta D_k^4 / 8\, (\varepsilon_k^L)^3)\{(\exp(\beta \varepsilon_k)+1)^{-1} - (\exp(\beta \varepsilon_{k+Q})+1)^{-1}\}$$

$$+ \sum_k (\beta^2 D_k^4 / 8\, (\varepsilon_k^L)^2)\{\exp(\beta \varepsilon_k)(\exp(\beta \varepsilon_k)+1)^{-2} + \exp(\beta \varepsilon_{k+Q})(\exp(\beta \varepsilon_{k+Q})+1)^{-2}\} \quad (19)$$

and

$$\sum_{k,r(=U,L),v} (\beta E_k^{(r,v)}) (\exp(\beta E_k^{(r,v)}) +1)^{-1} \approx \sum_{k,j=1,2} (\beta \varepsilon_j(k)) (\exp(\beta \varepsilon_j(k)) +1)^{-1}$$

$$+ \sum_k (\beta D_k^2/2\, \varepsilon_k^L)\{ (\exp(\beta\varepsilon_k)+1)^{-1} - (\exp(\beta\varepsilon_{k+Q})+1)^{-1}\}$$

$$- \sum_k (\beta D_k^4/8\, (\varepsilon_k^L)^3)\{ (\exp(\beta\varepsilon_k)+1)^{-1} - (\exp(\beta\varepsilon_{k+Q})+1)^{-1}\}. \quad (20)$$

Now upon ignoring the dependence of chemical potential on $\beta$, from Eqs. (19) and (20), we are able to see that $\Delta s \approx (2/N_s)\sum_{k,j=1,2} s_j(k)$, where $s_j(k) = (\beta^{*2}/32)\{D_k^4 (T<T^*)/ (\varepsilon_k^L)^2\}$ sech$^2(\beta^*\varepsilon_j(k)/2)$. The entropy density ($\sum_{,j=1,2} s_j(k)$) plot on the Brillouin zone is shown in Fig.4 assuming $(\chi_0/\Delta_0 (T<T^*))^2 = 0.0025$. The positive and singular contribution $\sim 5\times10^{11}$ (see Fig.6 where the contributions appear almost as delta-peaks ) arise from the eight points('Hot Spots'):$(k_xa, k_ya)\approx\{(\pm 0.7628, 2.4121), (\pm 0.7628, -2.4121), (\pm 2.4121, 0.7628),$ $(\pm 2.4121, -0.7628)\}$ quite close to the RBZ boundary $k_y \pm k_x = \pm\pi/a$ and the electron pockets. The points within the RBZ practically do not contribute. The integral $(2/N_s)\sum_{k,j=1,2} s_j(k)$, therefore, may be expressed as

$(2/N_s) \sum_{k,j=1,2} s_j(k)$

$\approx 5\times10^{11}(\beta^{*2}\Delta_0^4 (T<T^*)/1024 t^2) \int_{-\pi}^{\pi}(dx/2\pi) \int_{-\pi}^{\pi}(dy/2\pi) \sum_{j=\pm 1} \{ \partial(x+2.4121)\partial(y+0.7628j)$

$+ \partial(x-2.4121)\partial(y+0.7628j) + \partial(x+0.7628)\partial(y+2.4121j) + \partial(x-0.7628)$

$\times \partial(y+2.4121j)\}, \quad (21)$

which, in turn, is $4\times10^{12}(\beta^{*2}\Delta_0^4 (T<T^*)/1024 t^2) \sim 1.7\times 10^{10} > 0$. We conclude from our theoretical analysis that the pseudo-gap(PG) transition is a first order one and the entropy per unit cell increases in this hidden-order state. The increase is about $2.37\times 10^{-13}$ J-K$^{-1}$.

## 4. Discussion on anomalous Nernst effect

We now discuss the anomalous Nernst effect (ANE) to highlight the fact that the BCs correspond to magnetic field in the momentum space. In the presence of an external electric field **E** along y-direction, the transverse heat current $J_x$ in the x-direction is given by the relation[12] $J_x = (T\, S_{xy}\, E_y)$ where the coefficient $S_{xy}$ is defined below. The current does not require the presence of a magnetic field. The reason is that, in the presence of chirality, the carriers acquire an anomalous velocity **v** given by $\hbar \mathbf{v_j} = e\, \mathbf{E} \times \mathbf{\Omega^{(j)}(k)}$. Upon multiplying this velocity by the entropy density ($k_B s_j(k)$) calculated above, we obtain the coefficient $S_{xy}$ for the transverse heat current: $S_{xy}(T<T^*) = (e/(d\hbar))\alpha_{xy}(T<T^*)$, $\alpha_{xy}(T<T^*) = (1/(N_s\, a^2))\sum_{k,j=1,2} \Omega^{(j)}(k)\, k_B\, s_j(k)$, where d =1.17 nm( the distance between consecutive two-dimensional layers of YBCO). For the pure DDW order this effect cannot arise due to absence of chirality. In Fig.5 we have plotted $(1/(N_s\, a^2))\sum_{j=1,2} \Omega^{(j)}(k)\, s_j(k)$, which is a dimensionless quantity, on the Brillouin zone taking $(\chi_0/\Delta_0 (T<T^*))^2 = 0.0025$. The positive and singular contribution ∼ $8\times 10^6$ (see Fig.7 where the contributions appear almost as delta-peaks ) arise, once again, from the eight points: $(k_x a, k_y a) \approx \{(\pm 0.762806, 2.4121), (\pm 0.762806, -2.4121), (\pm 2.4121, 0.762806), (\pm 2.4121, -0.7628)\}$ quite close to the RBZ boundary and the electron pockets. The negative contributions ∼ $-7.7\times 10^6$ also arise (see Fig.8) from these points as the BC can be positive as well as negative; the points within the RBZ practically do not contribute. The total contribution, therefore, remains finite and less than a number ∼ $10^6$. One can carry out the momentum integration as in Eq.(21) and obtain $\alpha_{xy}(T<T^*)$. Its expected value $<<10^{-16}$ J-K$^{-1}$. We have, thus, been able to show graphically the possibility of anomalous Nernst effect ( the effect in the absence of magnetic field) in a YBCO sample with the origin being the so called 'Hot Spots' in the RBZ boundary. Whether such effects are experimentally detectable is a moot question. It may be mentioned that the experimentally observed Nernst signal, $S_{xy}\rho$, (where $\rho$ is the magneto-resistivity) in high-temperature superconductors such as LSCO and BSCCO [15] at the temperature close to T* is ~ 100 nV-K$^{-1}$.

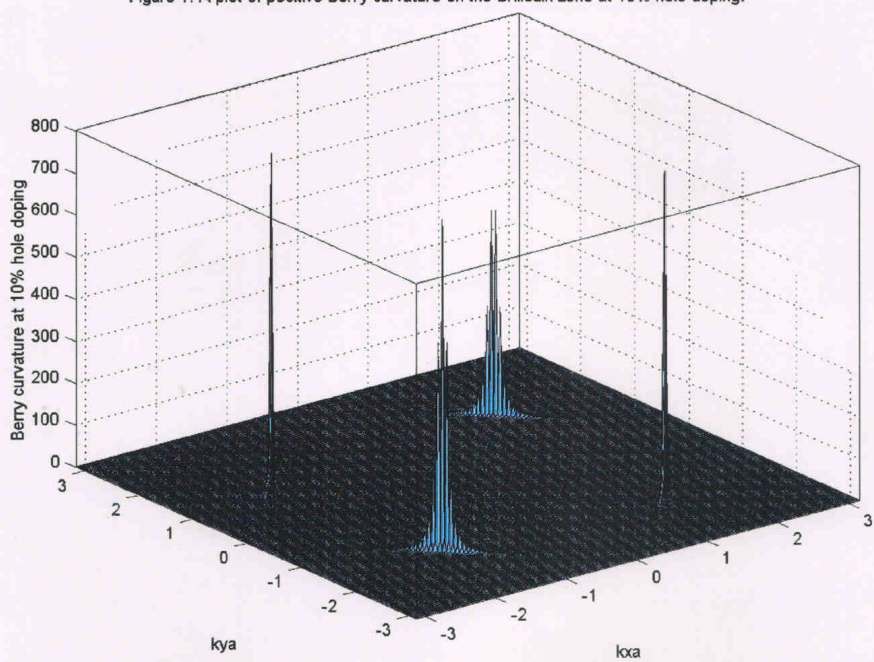

Figure 1: A plot of positive Berry curvature on the Brillouin zone at 10% hole doping.

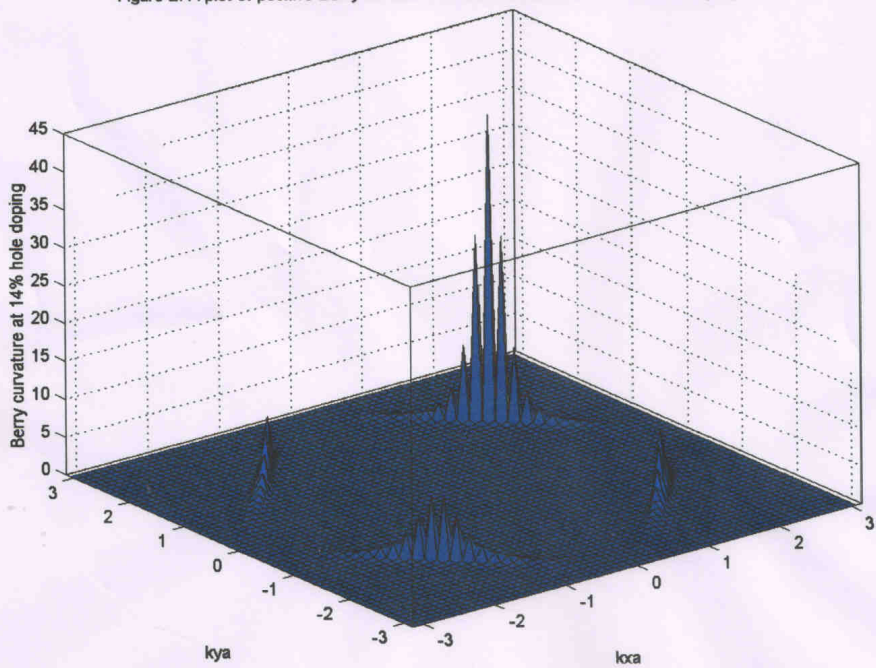

Figure 2: A plot of positive Berry curvature on the Brillouin zone at 14% hole doping.

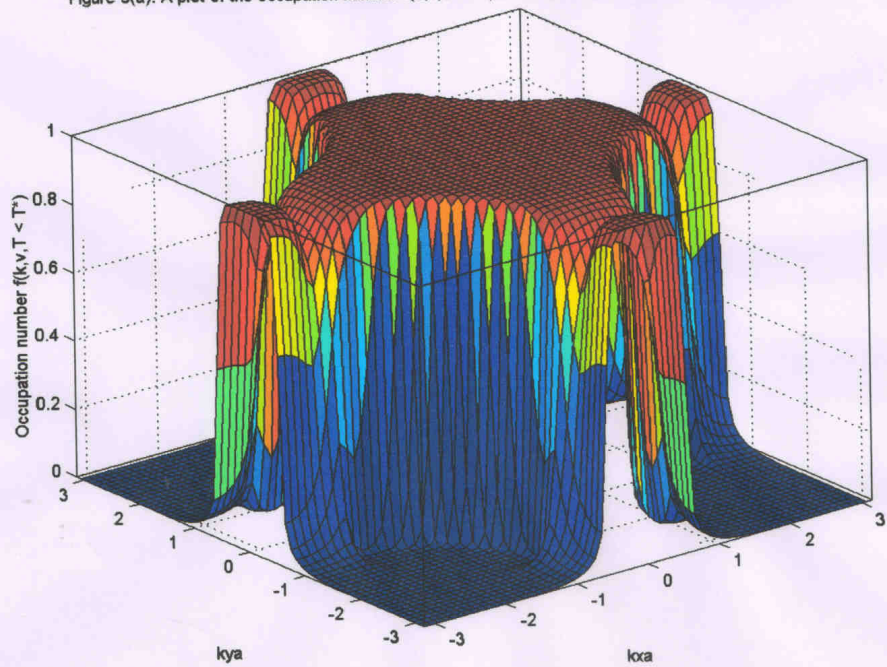

Figure 3(a): A plot of the occupation number $f(k,v,T < T^*)$ on the Brillouin zone at 10% hole doping.

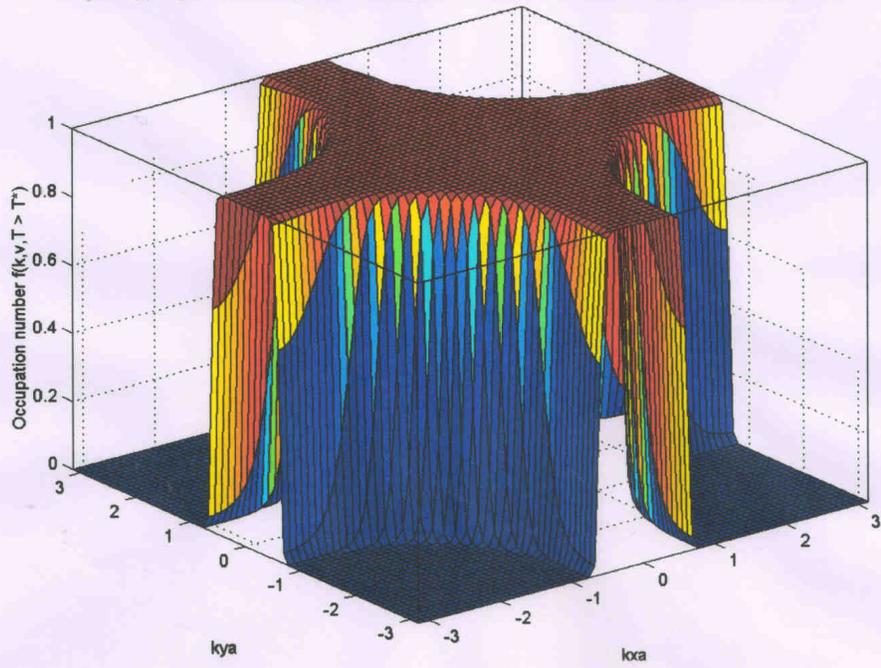

Figure 3(b): A plot of the occupation number $f(k,\nu,T > T^*)$ on the Brillouin zone at 10% hole doping.

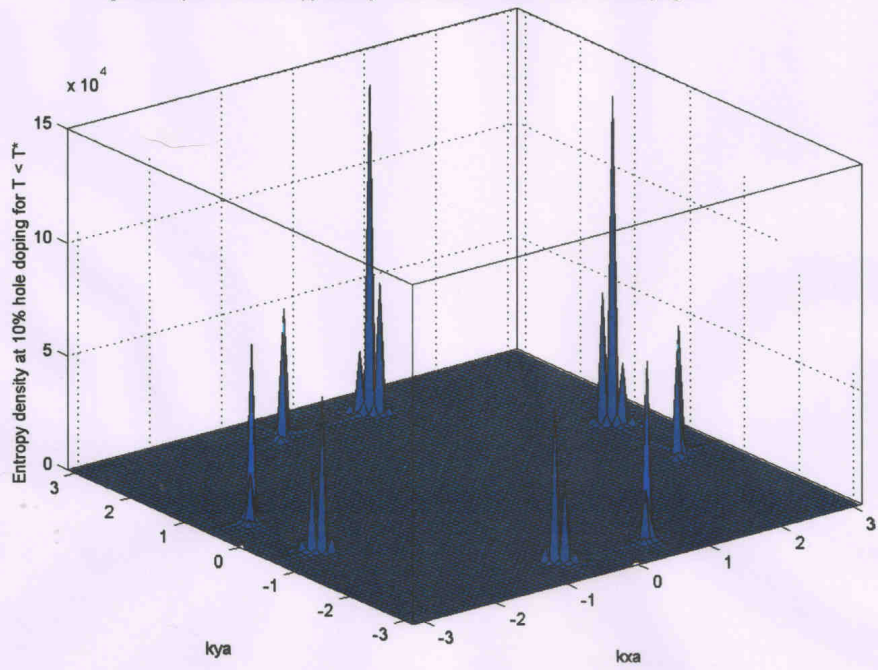

Figure 4: A plot of the entropy density on the Brillouin zone at 10% hole doping and T < T*.

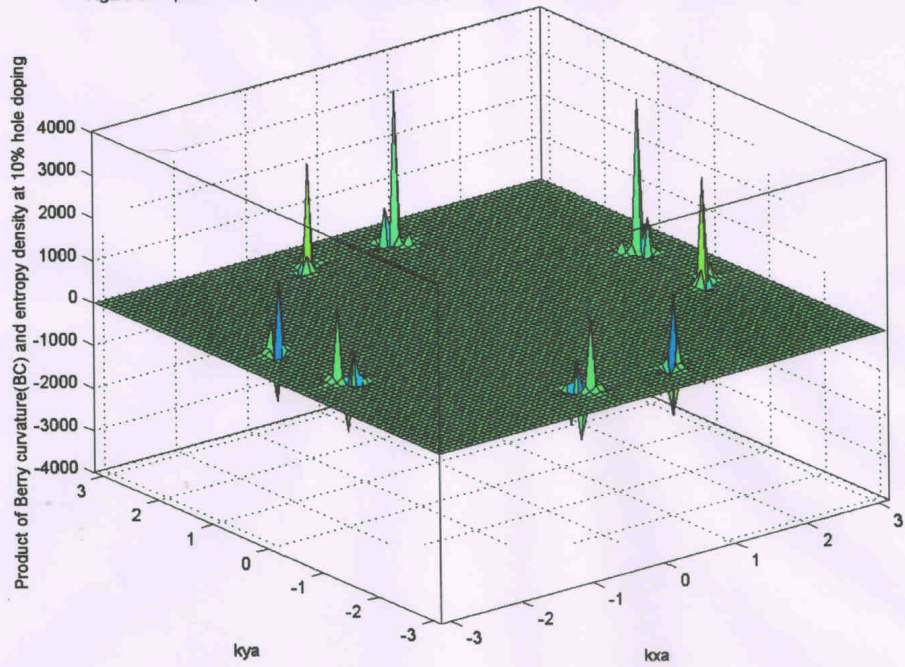

Figure 5: A plot of the product of BC and entropy density on the Brillouin zone at 10% hole doping.

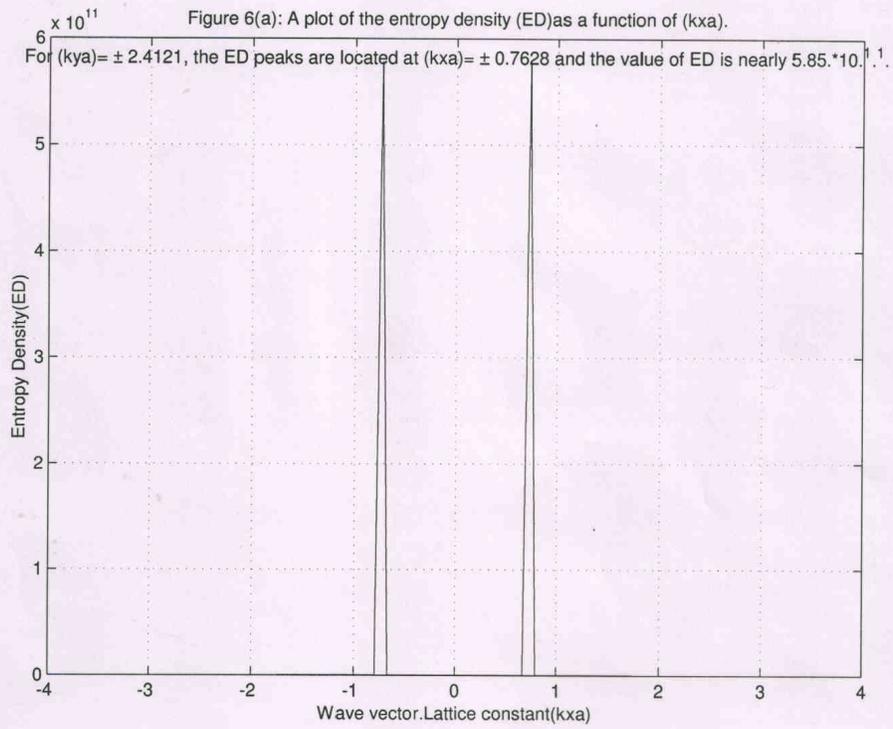

Figure 6(a): A plot of the entropy density (ED) as a function of (kxa). For (kya)= ± 2.4121, the ED peaks are located at (kxa)= ± 0.7628 and the value of ED is nearly $5.85 \times 10^{11}$.

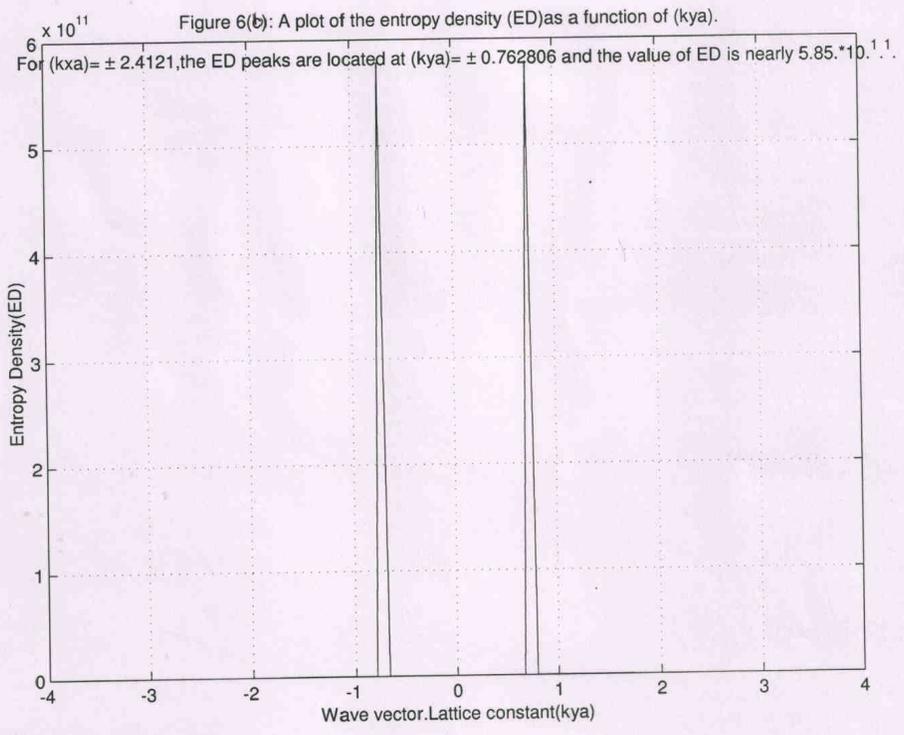

Figure 6(b): A plot of the entropy density (ED) as a function of (kya).
For (kxa)= ± 2.4121, the ED peaks are located at (kya)= ± 0.762806 and the value of ED is nearly $5.85 \times 10^{11}$.

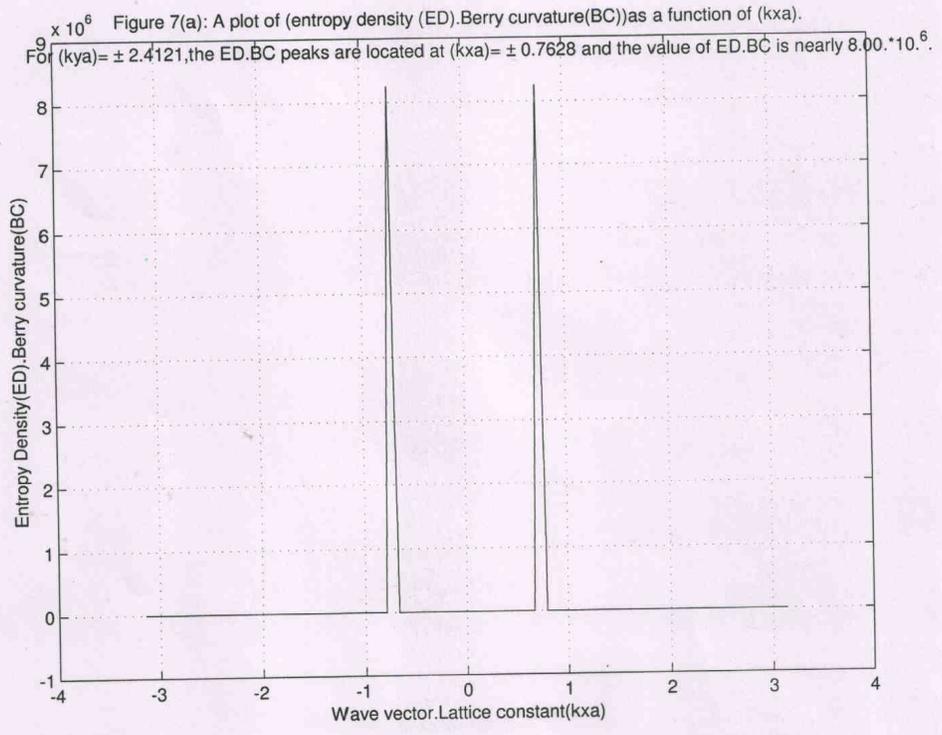

Figure 7(a): A plot of (entropy density (ED).Berry curvature(BC)) as a function of (kxa). For (kya)= ± 2.4121, the ED.BC peaks are located at (kxa)= ± 0.7628 and the value of ED.BC is nearly $8.00 \times 10^6$.

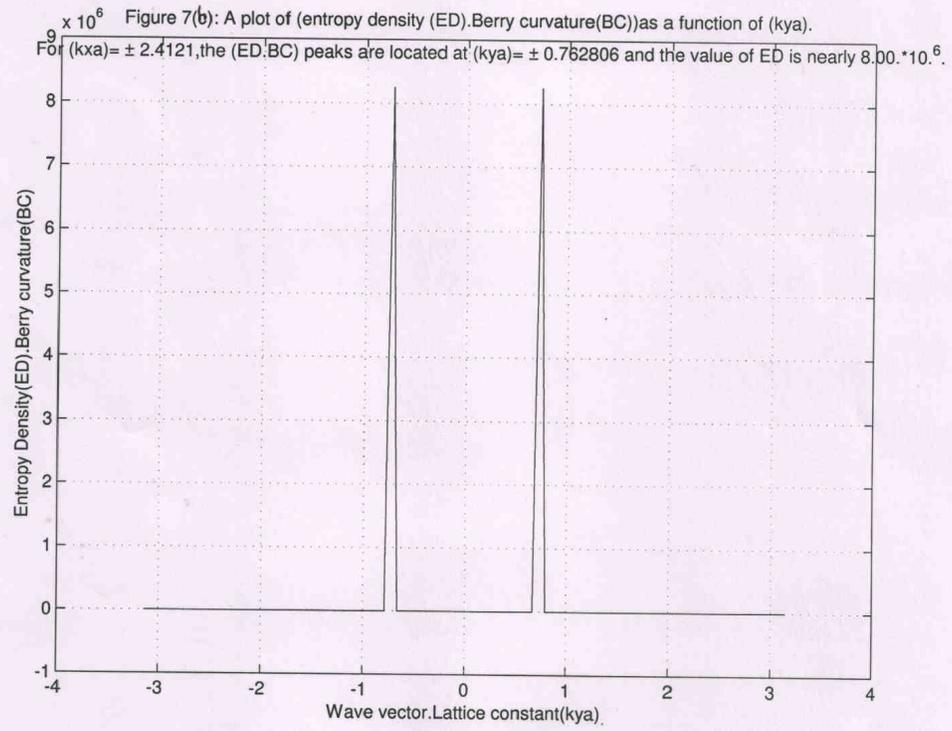

Figure 7(b): A plot of (entropy density (ED).Berry curvature(BC)) as a function of (kya). For (kxa)= ± 2.4121, the (ED.BC) peaks are located at (kya)= ± 0.762806 and the value of ED is nearly 8.00.*10$^6$.

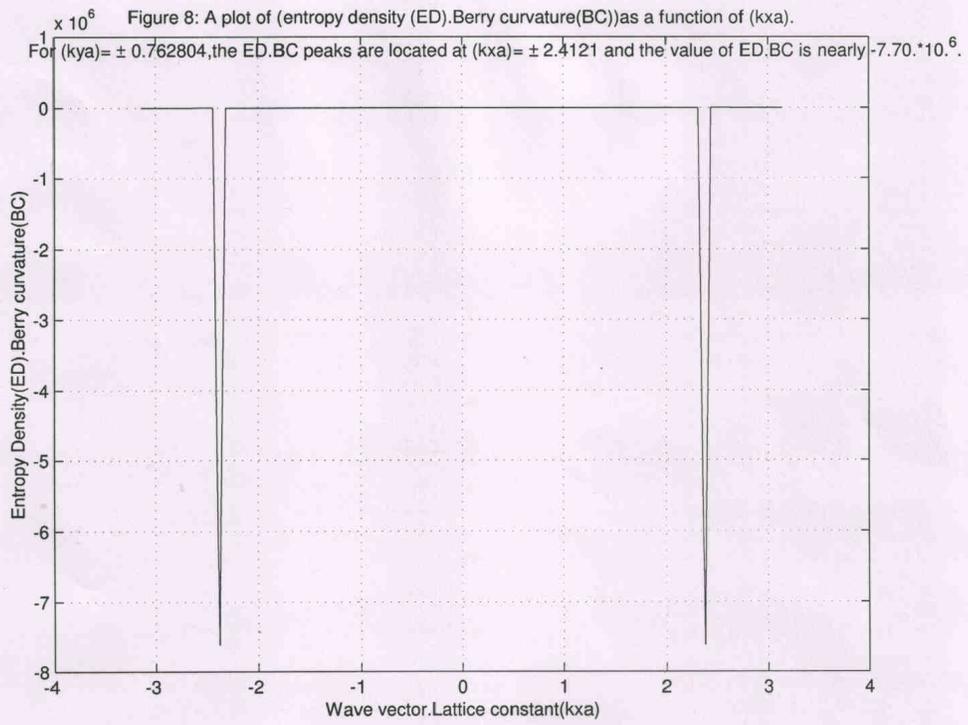

Figure 8: A plot of (entropy density (ED).Berry curvature(BC)) as a function of (kxa). For (kya)= ± 0.762804, the ED.BC peaks are located at (kxa)= ± 2.4121 and the value of ED.BC is nearly -7.70.*10.$^6$.